\documentclass[floatfix, noshowpacs, preprintnumbers, twocolumn, amsmath, amssymb, aps, prx, superscriptaddress]{revtex4-1}

\usepackage{graphicx, xcolor}
\usepackage{soul}
\usepackage{times}
\usepackage{multirow}
\usepackage{booktabs}
\usepackage{hhline}
\usepackage{rotating}
\usepackage{complexity}
\usepackage{color}
\usepackage{hyperref}
\usepackage{booktabs}
\usepackage{colortbl}
\usepackage{mathtools}
\usepackage[fleqn]{nccmath}

\definecolor{Gray}{gray}{0.9}

\DeclarePairedDelimiter{\ceil}{\lceil}{\rceil}

\begin{document}

\title{Photonic architecture for reinforcement learning}

\author{Fulvio Flamini}
\author{Arne Hamann}
\author{Sofi\`{e}ne Jerbi}
\author{Lea M. Trenkwalder}
\author{Hendrik Poulsen Nautrup}
\author{Hans J. Briegel}
\affiliation{Institut f\"{u}r Theoretische Physik, Universit\"{a}t Innsbruck, Technikerstra{\ss}e 25, 6020 Innsbruck, Austria}


\begin{abstract}
The last decade has seen an unprecedented growth in artificial intelligence and photonic technologies, both of which drive the limits of modern-day computing devices. In line with these recent developments, this work brings together the state of the art of both fields within the framework of reinforcement learning. We present the blueprint for a photonic implementation of an active learning machine incorporating contemporary algorithms such as SARSA, Q-learning, and projective simulation. We numerically investigate its performance within typical reinforcement learning environments, showing that realistic levels of experimental noise can be tolerated or even be beneficial for the learning process. Remarkably, the architecture itself enables mechanisms of abstraction and generalization, two features which are often considered key ingredients for artificial intelligence. The proposed architecture, based on single-photon evolution on a mesh of tunable beamsplitters, is simple, scalable, and a first integration in portable systems appears to be within the reach of near-term technology.
\end{abstract}

\maketitle

\section{Introduction}

Modern computing devices are rapidly evolving from handy resources to autonomous machines \cite{Iliadis18}. On the brink of this new technological revolution \cite{Schwab15}, reinforcement learning (RL) has emerged as a powerful and flexible tool to enable problem solving at an unprecedented scale \cite{Mnih2015, Silver2016, openAI, Silver18, Arulkumaran19}.
This breakthrough development was in part spurred by the technological achievements of the last decades, which unlocked vast amounts of data and computational power.
One of the key ingredients for this advancement was the ultra-large-scale integration \cite{Meindl84}, which led to the massive capabilities of current portable devices. 
Meanwhile, in the wake of this technological progress, neuromorphic engineering \cite{Thakur18}  was developed to mimic neuro-biological systems on application-specific integrated circuits (ASIC) \cite{Mead90}. 
Their improved performance is rooted in the parallelized operation and in the absence of a clear separation between memory and processing unit, which eliminates off-circuit data transfers.
Furthermore, new materials and ASICs are being reported to boost neuromorphic applications \cite{Islam19}. Among them, photonic devices represent a promising technological platform due to their fast switching time, high bandwidth and low crosstalks \cite{deLima19}.

Inspired by the outstanding success of both RL and ASICs, here we present a novel photonic architecture for the implementation of active learning agents. More specifically, we consider an RL approach to artificial intelligence \cite{Sutton98}, where an autonomous agent learns through interactions with an environment. Within this framework, the proposed architecture can operate using any of three learning models: SARSA \cite{Rummery1994}, Q-learning \cite{Watkins1989} and projective simulation (PS) \cite{briegel2012projective}.
The main contribution of this paper is twofold. (i) First, we describe a photonic architecture that enables RL algorithms to act directly within optical applications.
To this purpose, we focus on linear-optical circuits for their intuitive description, well-developed fabrication techniques and promising features as compared to electronic processors \cite{Sun15, Komljenovic16, Flamini_2018, Atabaki2018}. For instance, nanosecond-scale routing and reconfigurability have already been demonstrated \cite{Harris2018, Pérez2018, Stabile2016}, while encoding information in photons enables decision-making at the speed of light, only limited by the generation and detection rates.
Moreover, the use of phase-change materials for in-memory information processing \cite{Rios19} promises to enhance the energy efficiency, since their properties can be modified without continuous external intervention \cite{Wuttig17, Miller18}.
(ii) The second contribution is the development of a specific variant of PS based on binary decision trees (tree-PS, or t-PS for short), which is closely connected to the standard PS and suitable for the implementation on a photonic circuit. Furthermore, we discuss how this variant enables key features of artificial intelligence, namely abstraction and generalization \cite{Ponsen10, melnikov2017generalization}.

The paper is structured as follows. In Sec. \ref{sec:RL} we summarize the theoretical framework of RL, exemplified by three common approaches: SARSA, Q-learning, and PS.
In Sec. \ref{sec:pPS} we describe the blueprint for a  fully integrated, photonic RL agent. We then numerically investigate its performance within two standard RL tasks and under realistic experimental imperfections in Sec. \ref{sec:TestingGridWorld}. Finally, in Sec. \ref{sec:multilayerPS} we discuss promising features of this architecture within the context of t-PS.


\section {Reinforcement learning}
\label{sec:RL}

\begin{figure}[t!]
	\centering
	\includegraphics[width=1 \linewidth]{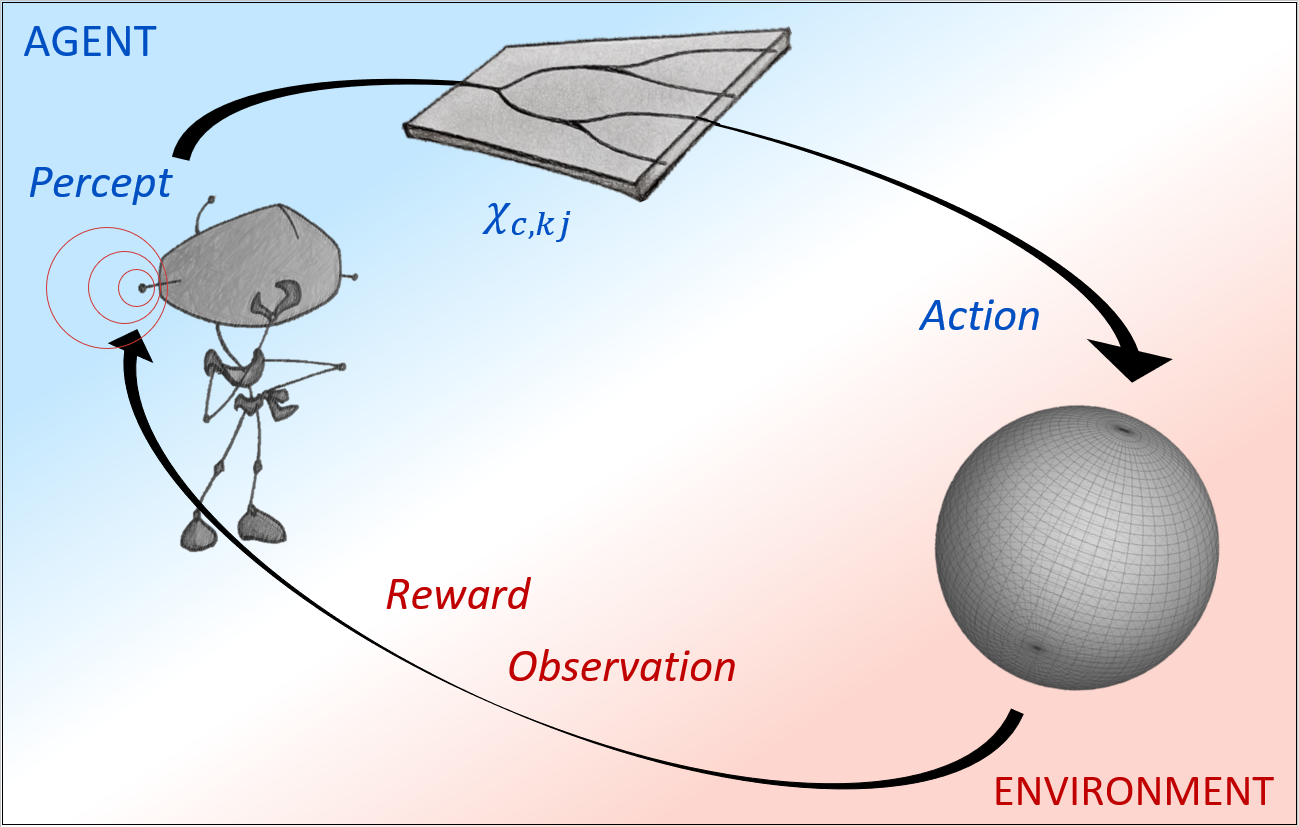}
\caption{\textbf{Reinforcement learning in a photonic circuit.} In RL, an agent learns by interacting with its environment. Each new observation is internally processed until an action is chosen and performed on the environment.
The processing unit, characterized by $\chi$-values, adapts the agent’s behavior according to a specific update rule in order to maximize the expected, future reward within a given environment. This unit can be implemented on an integrated photonic circuit. }
\label{fig:1}
\end{figure}

In this section, we briefly introduce the RL framework, which is the focus of this work. Within RL, the agent learns through a cyclic interaction with the environment (Fig. \ref{fig:1}). The agent starts with no prior knowledge and randomly probes the environment by performing actions. The environment, in turn, responds to the actions by changing its state, which is observed by the agent through perceptual input, and by providing a reward that quantifies how well the agent is performing. The goal of the agent is then to maximize its long-term expected reward \cite{SUTTON90}.
In the following, we will first describe two standard RL algorithms, SARSA \cite{Rummery1994} and Q-learning \cite{Watkins1989}, before introducing the more recent PS \cite{briegel2012projective}.

\subsection{SARSA and Q-learning}
\label{sec:S_QL}

As for all RL algorithms, SARSA (State-Action-Reward-State-Action) and Q-learning aim at adjusting the agent's behavior until it performs optimally, in the sense we discuss in the following. The agent's behavior is defined by the policy $\pi_{a|s}$, which governs the choice of an action $a \in A$ given a state $s \in S$. The evolution of the environment under the agent's action can be described by a conditional probability distribution over all state-action-state transitions. Each transition that was taken has an associated reward $\lambda$. For a given policy $\pi_{a|s}$, the value of each state is defined by the expected future reward $V^{\pi}_s =\mathbb{E}[\sum_{t=0}^{T}\gamma^{t} \lambda_t]$. Here, $\lambda_t$ is the reward received from the environment at time $t$, while the so-called discount factor $\gamma \in [0,1]$ sets the relative importance of immediate rewards over delayed rewards, up to a temporal horizon $T$.
The goal of the agent is to learn the optimal policy $\pi^*_{a|s}$ that maximizes the value $V^{\pi}_s$ for all states $s$. 
The expected future reward is estimated and iteratively updated through the experience gained from its interactions with the environment.
Instead of the value $V^{\pi}_s$, this estimate is more conveniently described by the $Q$-value, which quantifies the quality of a state-action pair at a given time (Fig. \ref{fig:2}a). For both SARSA and Q-learning, this quantity is updated at each step according to
\begin{equation}
Q^{(t+1)}_{s,a}=(1-\alpha) \ Q^{(t)}_{s,a} + \alpha \ \tilde{Q}^{(t)}_{s',a'}
\label{eq:update_SandQL}
\end{equation}
where $\tilde{Q}^{(t)}_{s^{\prime}, a^{\prime}} = \lambda_{s^{\prime}, a^{\prime}} + \gamma  f ( Q^{(t)}_{s^{\prime}, a^{\prime}} )$ is the new estimate due to taking action $a'$ in the state $s'$ observed after $s$, $f$ is a suitable function that depends on the algorithm and the learning rate $\alpha$ determines to what extent this estimate overrides the old value. Given $N$ actions and as many $Q$-values for state $s$, the expected future reward can be estimated as $V_s=\sum_{j=1}^N \pi_{j|s}  Q_{s,j}$.

In both algorithms, decision-making is usually done by sampling actions according to a probability distribution that depends on the $Q$-values. In the context of RL, the softmax function is a convenient choice  
\begin{equation}
   \pi_{a|s}(Q_{s,a})  = \frac{\mathrm{e}^{\beta Q_{s,a} }}{\sum_{j=1}^{N} \mathrm{e}^{\beta Q_{s, a_j}}}
\label{eq:PS_prob_softmax} 
\end{equation}
\noindent where the parameter $\beta$ governs the drive for exploration within the agent.
The difference between Q-learning and SARSA lies in the choice of the function $f$. In SARSA, $f$ updates the value of the current state-action pair $Q_{s,a}$ with the estimate for the following state-action pair $Q_{s',a'}$, i.e. $f$ is the identity function. In state $s'$, the action $a'$ is chosen according to the agent's policy. Thus, SARSA is called an on-policy algorithm. Q-learning, on the other hand, is an off-policy algorithm because, given the state $s'$, $f$ selects the action $a'$ with the maximal value $Q_{s',a'}$, i.e. $f=\max_{a'\in A}$, so that the update is independent of the next action chosen according to the agent's policy.

\subsection{Projective simulation}
\label{sec:PS}

\begin{figure*}[t!]
	\includegraphics[width=0.875\linewidth]{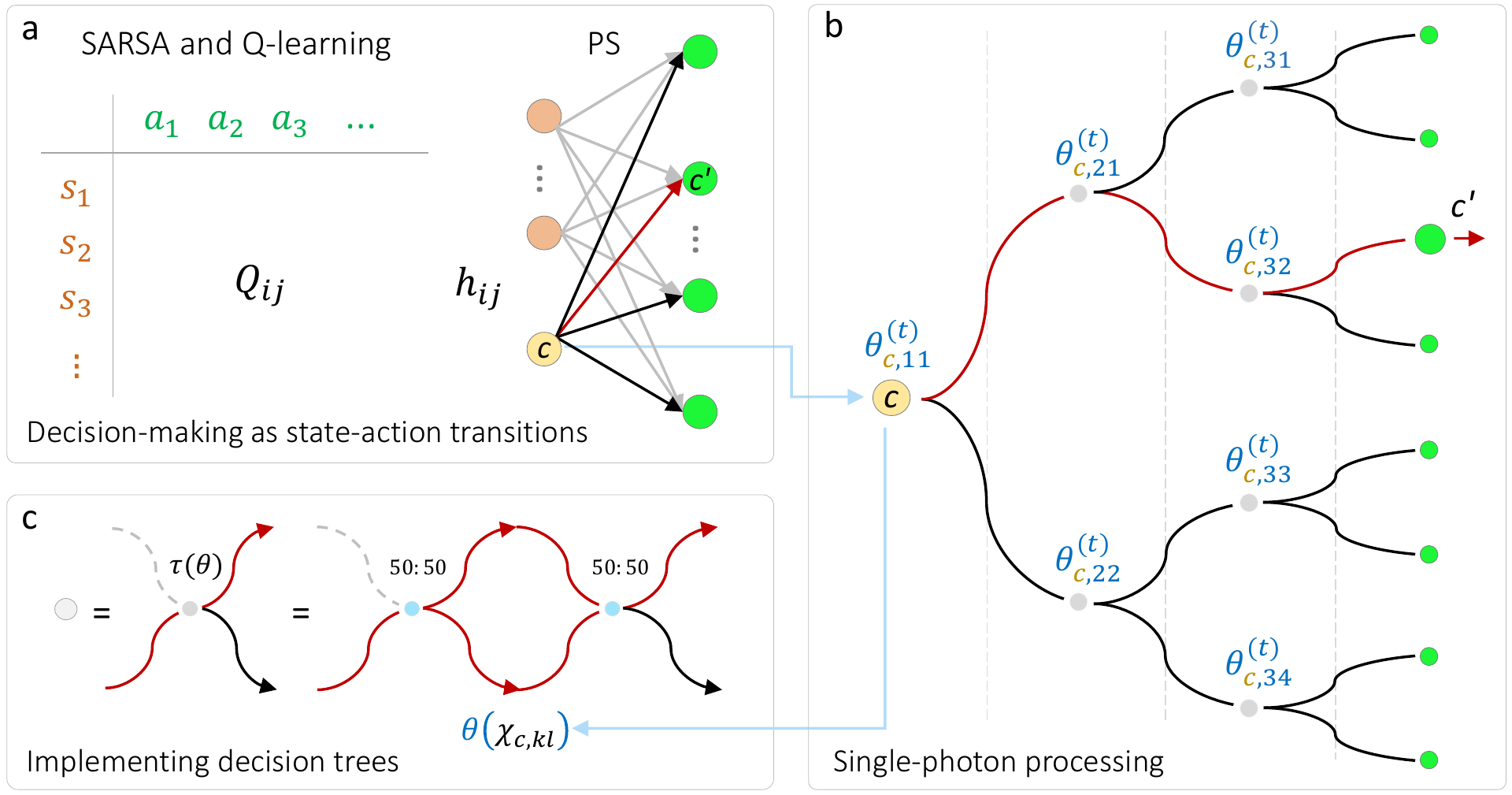}
	\caption{\textbf{Photonic implementation of reinforcement learning.} a) Learning and decision-making are decomposed into sequences of state-action (SARSA and Q-learning) or clip-to-clip (PS) transitions. Every time a clip ($c$) is hit, a new clip ($c'$) is chosen according to the model and the current policy. For a two-layer ECM, clip-to-clip transitions are equivalent to state-action pairs. Learning develops by updating suitable quantities: $Q$-values (SARSA and Q-learning), $h$-values (PS) and $\chi$-values for their photonic implementation.
	b) Arbitrary probabilistic transitions can be implemented on a photonic platform with a cascade of beamsplitters, whose transmissivities $\tau$ reproduce the distribution given by the current policy. c) Tunable beamsplitters at each node $(k,l)$ can be implemented using Mach-Zehnder interferometers with tunable phase shifts $\theta_{c,kl}$ \cite{Miller13self_config}. Phases are adjusted according to a quantity $\chi_{c,kl}$, which is updated during the learning process.}
	\label{fig:2}
\end{figure*}

PS is a recent, physically-motivated RL model \cite{briegel2012projective}, which has already found several applications ranging from robotics \cite{simon2016} and quantum error correction \cite{poulsennautrup2018optimizing} to the study of collective behavior \cite{Ried19} and automated experiment design \cite{melnikov2017active}. 
Decision-making in PS occurs in a network of clips that constitutes the agent's episodic and compositional memory (ECM) (Fig. \ref{fig:2}a). Each clip represents a remembered percept, a remembered action or a more complex combination thereof. The ECM can accommodate a multilayer structure, where intermediate layers represent abstract clips and connections. Decision-making is carried out by a random walk through the ECM, starting at a percept clip and ending at an action clip which triggers the corresponding action. The random walk is guided by transition probabilities between pairs of clips $(c_i,c_j)$,  connected by edges carrying weights $h_{c_i,c_j}$,
by considering probabilities proportional to $h_{c_i,c_j}$ 
or by using the softmax function $\pi_{c_j|c_i}(h_{c_i,c_j}) $ as in Eq. \ref{eq:PS_prob_softmax}.
Learning occurs by updating the clip network in the agent's memory, i.e. by changing its topology or the edge weights $h_{c_i,c_j}$. In the latter case, the update rule at time $t$ has the form
\begin{equation}
h^{(t+1)}_{c_{i}, c_{j}}=h^{(t)}_{c_{i}, c_{j}} -\gamma\left(h^{(t)}_{c_{i}, c_{j}}-1\right) +g_{c_{i}, c_{j}} \lambda
\label{eq:update_h}
\end{equation}
\noindent where $\gamma \in [0,1)$ is a damping parameter, $\lambda$ is the reward and  $g_{c_{i}, c_{j}} \in [0,1]$ is the so-called edge-glow value or $g$-value. Here, $\gamma$ and $g_{c_i,c_j}$ implement mechanisms that take into account forgetting and delayed rewards, respectively. 
More specifically, the damping parameter $\gamma$ is essential for environments that change over time, effectively damping $h$-values at each time step. The edge-glow values serve to backpropagate discounted rewards to earlier sequences of actions. The $g$-values are updated at each time step: whenever an edge $(c_i,c_j)$ is traversed $g_{c_i,c_j}$ is set to $1$, and from then on its value is discounted as $ g^{(t+1)}_{c_i,c_j} =  (1-\eta) \, g^{(t)}_{c_i,c_j} $ where $\eta \in [0,1]$ is the glow parameter.
%
%
Consequently, $g$-values are rescaled according to $g_{c_i,c_j}=(1-\eta)^{\delta t_{c_i,c_j}}$, where $\delta t_{c_i,c_j}$ is the number of steps between the round when $(c_i,c_j)$  is traversed and the round when a reward is issued. 
Intuitively, values of $\eta$ close to 1 reward sequences of actions only in the immediate past, while values close to 0 are used to reward longer sequences. The glow parameter is relevant in environments with delayed rewards such as the GridWorld \cite{Sutton98} discussed in Sec. \ref{sec:TestingGridWorld}. 
For a more detailed description of PS we refer the reader to Refs. \cite{melnikov2017generalization, melnikov2018benchmarking, makmal2016meta}.


\section{Photonic reinforcement learning}

\label{sec:pPS}

In order to implement RL on a photonic platform we need to be able to satisfy two requirements: (i) implement arbitrary probabilistic transitions between clips and (ii) update the corresponding probability distributions in a controlled and effective way. A practical platform has to satisfy further criteria that are crucial for any implementation, such as scalability, ease of fabrication and miniaturizability.
In this section, we will describe a linear optical architecture that is tailored to the task at hand, i.e. designing integrated photonic hardware for RL, in the spirit of neuromorphic engineering \cite{deLima19}.

\subsection{Decision trees as linear optical circuits}

\label{sec:photonic_multiPS}

Using a bottom-up approach \cite{Note1}
, we focus on the implementation of state-action (SARSA and Q-learning) or clip-to-clip (PS) transitions, as shown in Fig. \ref{fig:2}a. For PS, each clip-to-clip transition is a building block for the random walk in the agent's memory.
For brevity, we will only consider clip-to-clip transitions ($c$,$c'$), which are equivalent to state-action pairs ($s$, $a$) for a two-layer ECM.
Each transition is governed by the probability distribution of detecting a single photon over the output modes. The architecture we present consists of a cascade of reconfigurable beamsplitters arranged in a tree structure (Fig. \ref{fig:2}b), which maps a single input mode (associated with a clip) to $N$ output modes (corresponding to as many clips). Such an association can be initialized randomly or according to prior knowledge about the environment.
Fully-reconfigurable linear-optical interferometers like this one allow to engineer arbitrary probability distributions over the optical modes and, given a probability distribution, it is possible to determine a set of phases that reproduces it exactly (see Sec. \ref{sec:programming_binary_tree}). In the next section, we also provide further considerations on various layouts that can be adopted.

To employ this architecture for RL, we consider the following operational scenario: the current policy is stored electronically \cite{Komljenovic16, Atabaki2018} in the phase shifters that define the single-photon evolution in the circuit and, consequently, the probabilistic decision-making. Each phase-shifter $\theta_{c,kl} $ at node ($k$,$l$) is set to implement the transition probabilities for the corresponding clip-to-clip connections. Decision-making (Fig. \ref{fig:2}a) is hence realized as a single-photon evolution in a mesh of tunable beamsplitters (Fig. \ref{fig:2}b), where the transition to the next state is made by detecting \cite{Note2}
Overall, this approach satisfies the requisite for arbitrary probabilistic transitions (i) described at the beginning of this section. Furthermore, it provides a solution that is scalable (one only needs to store the phases that implement a given transition) and that can be fully integrated on a miniaturized photonic chip \cite{Atabaki2018}.
Importantly, sensors could be integrated on an optical chip, gyroscopes and magnetometers being first examples in this direction \cite{Komljenovic16}.

Concerning the second requirement (ii), to learn an optimal policy we want the agent to autonomously adjust the phases $\theta$ according to a suitable update rule. To this end, we first consider the path $\Gamma_{c,c'}$ that connects clips ($c$,$c'$) and express the phases $\theta_{c,kl}$ in the transition probability 
\begin{equation}
p_{c'|c} = \prod\limits_{(k,l) \in \Gamma_{c,c'}} \sin^2 \theta_{c,kl}
\label{eq:transition_prob}
\end{equation}
\noindent as a function of a quantity $\chi$ that is updated during the learning process, namely $\theta (\chi) = \theta_0 + \theta_{\chi}$. Here, $\theta_0 = \frac{\pi}{4}$ corresponds to the configuration where all transitions are equally probable, while $\theta_{\chi}$ spans the whole range of transition probabilities, namely  $\theta_{\chi}\in [-\frac{\pi}{4},\frac{\pi}{4}]$. Suitable candidates for $\theta_{\chi}$ are the sigmoid functions \cite{sigmoid}, which are monotonically increasing in a bounded interval and have domain over all real numbers.
We then use the function $\theta_{\chi}=\tanh \chi$, so that
\begin{equation}
\theta (\chi) = \frac{\pi}{4} (1 +\tanh \chi) 
\label{eq:photonic_map}
\end{equation}
\noindent where the quantity $\chi$ is updated according to  a suitable update rule within the framework of RL. For SARSA (S) and Q-learning (QL), we update $\chi$ according to the rules
\begin{equation}
    \begin{array}{c}
		\chi_{c,kl}^{S} \leftarrow \, (1-\alpha) \ \chi_{c,kl}^{S} + \alpha \left( \lambda_{c} + \gamma R_{c'} \right)   \\
		\ \\
		\chi_{c,kl}^{QL} \leftarrow \, (1-\alpha) \ \chi_{c,kl}^{QL} + \alpha \left( \lambda_{c} + \gamma R_{c'} M_{c'}  \right)  \\
    \end{array}
\label{eq:update_chi_SandQL}
\end{equation}
\noindent where $ M_{c'} = \max\limits_{c''} \big( \tanh  \sum\limits_{(k,l) \in \Gamma_{c',c''}} \frac{ | \chi_{c',kl}| }{n}  \big) $ , $n= \ceil{\log_2N}$ being the depth of the circuit, and
\begin{equation}
R_{c} \leftarrow \, (1-\alpha) \ R_{c} + \alpha \left( \lambda_{c} + \gamma R_{c'}  \right)
\label{eq:update_R}
\end{equation}
In the notation used in Sec. \ref{sec:S_QL} for SARSA and Q-learning, subscripts in $R_{c}$ and $ M_{c'}$ refer to states. Comparing the original Q-value update rule in Eq. \ref{eq:update_SandQL} with the update rule in Eq. \ref{eq:update_chi_SandQL}, we emphasize that Eq. \ref{eq:update_chi_SandQL} does not simply reproduce Eq. \ref{eq:update_SandQL} using $\chi$. The reason is that $Q$- and $\chi$-values provide different information, the former quantifying the quality of a clip-to-clip connection, the latter defining the splitting ratio at each beamsplitter. Indeed, though related once the agent has properly learned the policy, the two quantities are not directly linked during the learning process.
For instance, when one clip-to-clip connection ($c$,$c'$) is favorable (large $Q$-value) the policy $\pi_c$ is peaked (i.e. $\chi$-values far from zero), but when multiple ($c$,$c'$) pairs are favorable (large $Q$-values) the policy $\pi_c$ is less peaked ($\chi$-values closer to zero).
Therefore, a feature we demand is to keep track of the overall quality of each state, from which the $\chi$-values will reproduce the relative quality of each ($c$,$c'$) connection. We fulfill this task in Eq. \ref{eq:update_R}, introducing a new parameter (in addition to the $N-1$ phases) that updates the agent's confidence in the quality of clip $c$. Also, peakedness of each policy can be quantified by the average deviation from 0 (corresponding to a flat distribution) of the $\chi$s in each path, as done in $ M_{c'}$ in Eq. \ref{eq:update_chi_SandQL}. 

Besides SARSA and Q-learning, we can choose to operate in the framework of PS. In this case, we  evolve $\chi$ according to the rule
\begin{equation}
\chi_{c,kl}^{PS}  \leftarrow \gamma \, \chi_{c,kl}^{PS} + g_{c,kl}  \ \lambda_{c}
\label{eq:update_chi_PS}
\end{equation}
\noindent which is equivalent to the update rule for $h_{c,c'}$ in Eq. \ref{eq:update_h}, considering that $h_{c,c'}$ ($\chi_{c,kl}$) is initialized to 1 (0).
Notably, the choice $\theta_{\chi} = \tanh \chi$ in Eq. \ref{eq:photonic_map} establishes a formal connection between the proposed architecture and a specific variant of PS, which we call tree-PS (t-PS). This connection is derived in Sec. \ref{sec:multilayerSoftmax_update}. In t-PS, every clip-to-clip transition is implemented as a binary decision tree between the input and the output clips.
In Sec. \ref{sec:2Lstandard} and Sec. \ref{sec:2Lsoftmax} we prove that t-PS can reproduce the operation of the two-layer PS, which has been discussed extensively in the literature. While the two models appear to have the same representational power, t-PS provides an additional structure that can be exploited to enhance the learning process, as we describe below in Sec. \ref{sec:multilayerPS_generalization}.

\subsection{Photonic architecture for the agent's memory}
\label{sec:neuromorph_PS}

\begin{figure}[t]
	\includegraphics[width=\linewidth]{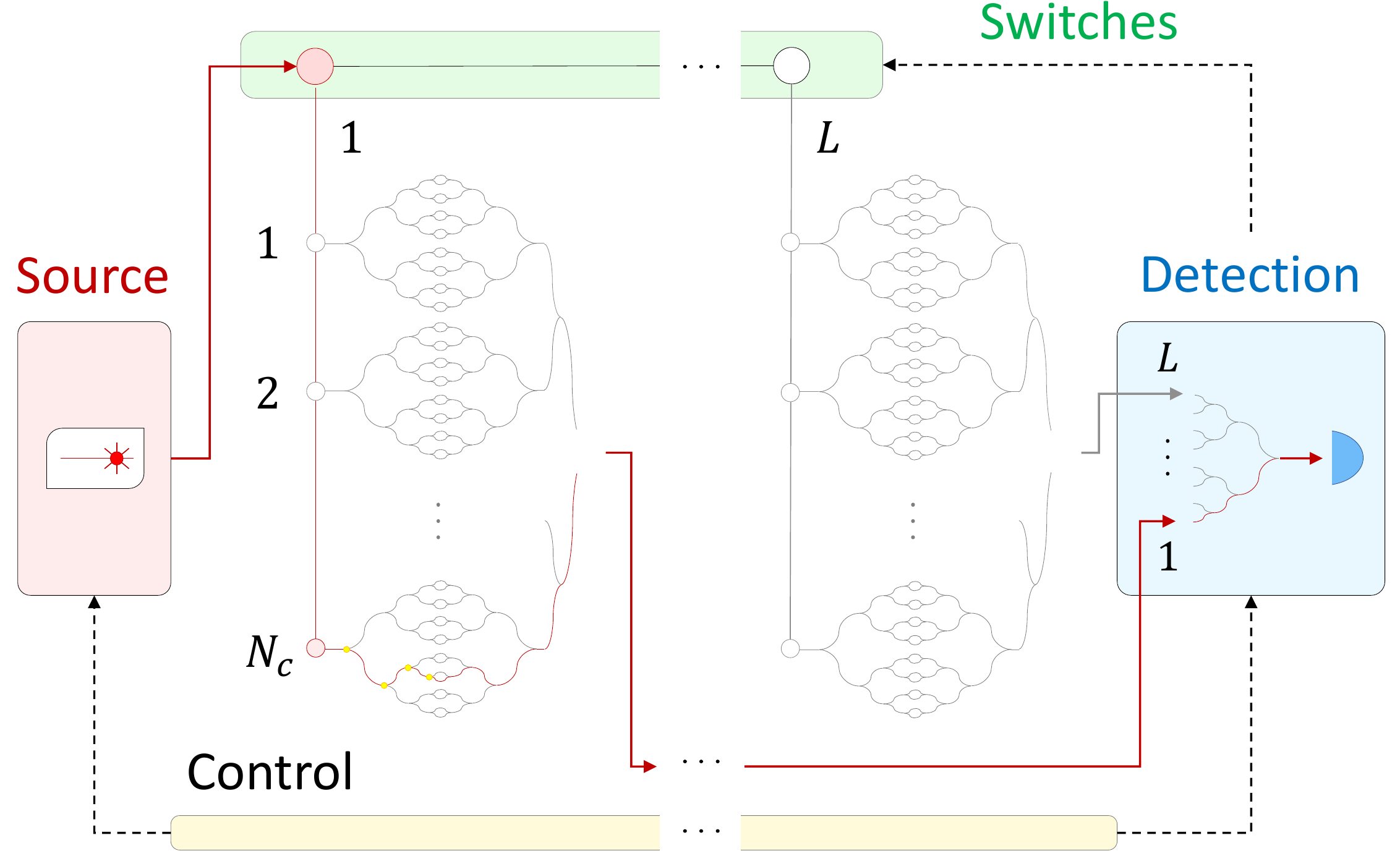}
	\caption{\textbf{Photonic architecture for learning and decision-making.} Single photons are routed by optical switches through $L$ layers, where state-action ($L=1$) or clip-to-clip transitions are performed (Fig. \ref{fig:2}).  For PS, in this paper we only consider acyclic ECMs. Photons are then time-multiplexed, using delay lines, before reaching the detection stage in a single waveguide (whose outcome controls the optical switches and possible updates). In principle the system can even be self-stabilized \cite{Miller13self_config, Miller13self_align, Grillanda14}.}
	\label{fig:3}
\end{figure}

The architecture described in Fig. \ref{fig:2}b, which represents the building block for decision-making, can take advantage of an efficient design enabled by its fractal geometry \cite{Taylor12}. In this section, we will outline three approaches to implement learning and decision making starting from such a building block. First, we can adopt a simple strategy where the circuit consists of a single decision tree: once a photon is detected (thus selecting a clip in the next layer), all phase-shifters are adjusted to implement the next transition and another photon is injected into the same circuit.
Similarly, we can devise a loop-based implementation where photons are redirected back to the input while the circuit is reconfigured. Though appealing, this approach is more challenging since it requires non-linearities to detect the presence of a photon in the output modes \cite{Imoto85}. Finally, we can conceive a more sophisticated scheme that fully exploits the advantages of a photonic platform. Here, all building blocks are arranged in a planar structure (Fig. \ref{fig:3}) that represents the memory of the agent (Fig. \ref{fig:2}a).
%
%
In the latter configuration, decision-making corresponds to a single-photon random walk from the input to the output layer. Input photons are routed through a bus waveguide and optical switches \cite{Miller13self_config, Tu19} to one layer (out of $L$), where a clip-to-clip transition is performed in a decision tree. The layered architecture is meaningful only for PS, where it represents the $L$-layer structure of an acyclic ECM, while for SARSA and Q-learning it is a convenient geometry to make the integrated circuit more compact.  Fast and efficient routing \cite{Nikolova15, Stabile2016}, controlled by a feedback system that also monitors photon losses, guides single photons to the appropriate building block. Photons exit the tree in one of $N$ waveguides (forming a second, reversed binary tree \cite{Wang18science}), whose root node leads to a second bus waveguide connected to the detection stage. To find out which clip (i.e. output waveguide) was selected, a possibility is to add $N-1$ different delay lines \cite{Zhuang07, Zhou18} to the reversed tree and look at the time bin where the photon was detected. 

An interesting feature of this approach is that it can take advantage of phase-change materials (PCM) \cite{Wuttig17, Miller18} to realize the phase-shifters,  whose physical properties can be modified in a reversible and controlled way with a single write operation \cite{Rios19}. The intuition is that only the phases corresponding to traversed paths need to be updated, while the others remain fixed without any additional power consumption. Hence, the number of updates scales only logarithmically with the number of output clips.
In Sec. \ref{sec_scaling}, 
we discuss how both computational complexity and energy consumption are even comparable to an electronic ASIC that exploits high locality and specialized data structure.
Notably, using the circuit for self-optimization in optical interferometers eliminates the need for a separate generation and detection, since photons can be part of the embedding application. In addition, decision-making after learning consumes practically no power since phase-shifters do not need to be adjusted anymore.



\section{Testing the architecture}

\label{sec:TestingGridWorld}

In this section, we employ the proposed architecture in a standard testbed for RL, the GridWorld environment \cite{SUTTON90}. This task is of broad relevance since any stationary fully-observable environment can be reformulated in this frame \cite{SUTTON90}, notable examples being Atari games \cite{Mnih2015} and Super Mario Bros. \cite{Togelius2009}. Henceforth, we will focus on (two-layer) PS, due to its simpler update rule (Eq. \ref{eq:update_chi_PS}) and to investigate the potential of t-PS.
Indeed, GridWorld has been already investigated in the context of PS \cite{melnikov2018benchmarking}, a relevant example being the design of  optical experiments, which was shown to be representable as a generalized GridWorld \cite{melnikov2017active}.  Furthermore, note that for both SARSA and Q-learning we numerically observed a performance very similar to PS.

In the simplest formulation of the problem, the goal for the agent is to maximize its long-term expected reward while navigating an environment structured as a planar grid-like maze. The agent starts from a fixed location $\vec{p}_A=(x_A, y_A)$  and is challenged to learn the shortest path that leads to a reward at location $\vec{p}_R=(x_R, y_R)$. Available to the agent is a set of actions $(x^\pm, y^\pm)$, where $x^{\pm}$ corresponds to a movement in the positive/negative $x$-direction.
The learning process is divided in a sequence of episodes, or trials, where the agent interacts with the environment until a predetermined condition is met. In our analyses, the agent is reset if the number of interactions in one episode either exceeds $10^3$ or a reward is obtained.
To account for delayed rewards, the edge-glow mechanism  (see Sec. \ref{sec:PS}) rescales the reward $\lambda$, assigned to a traversed transition ($c_i$,$c_j$), by a quantity that decreases exponentially with the number of steps that pass until a reward is received \cite{SUTTON90, melnikov2018benchmarking}.

\begin{figure}[h]
	\includegraphics[width=\linewidth]{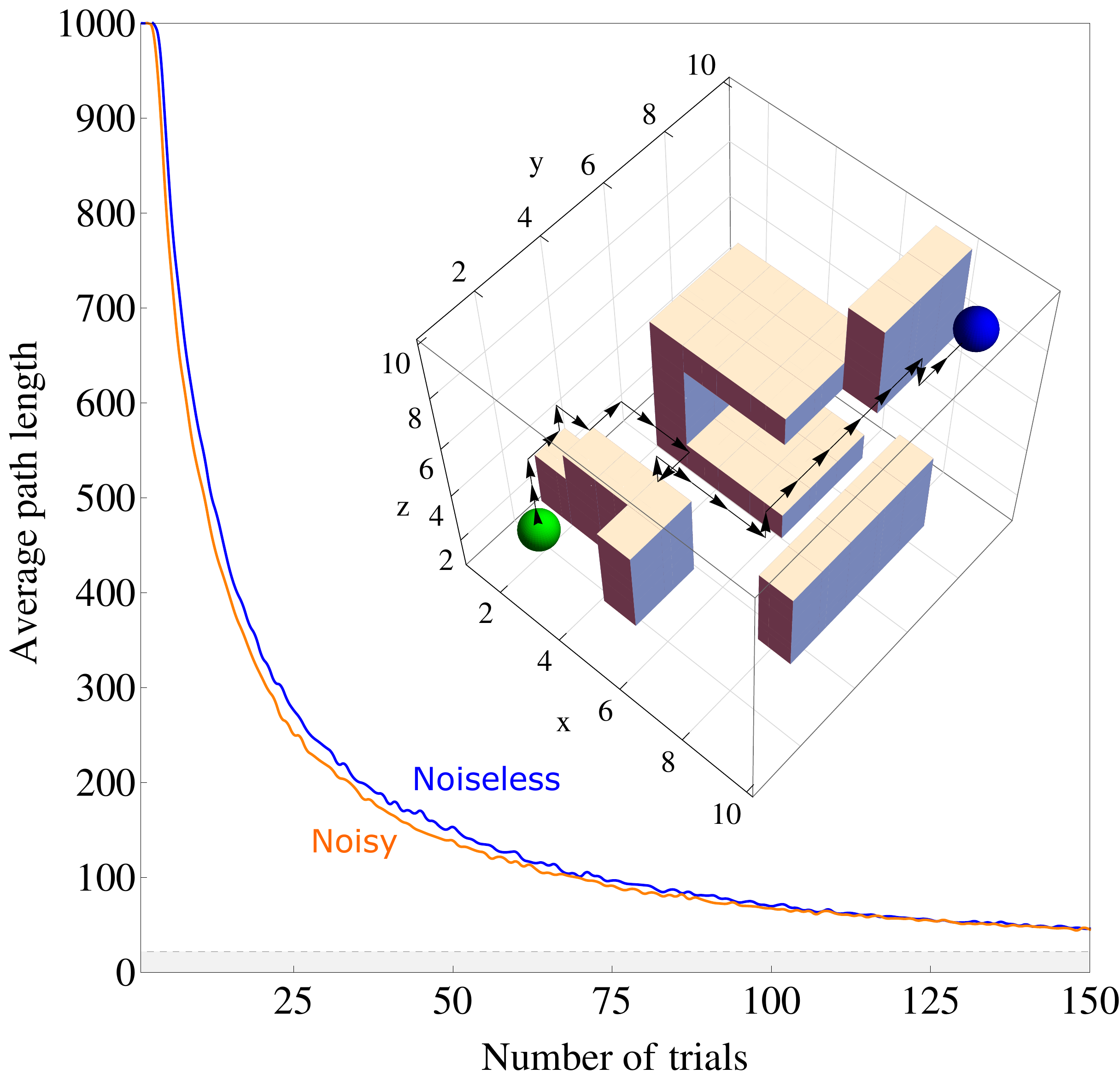}
	\caption{\textbf{Simulating the photonic architecture in GridWorld.}
	Average path length required by a PS agent to reach the reward in a $10 \times 10 \times 10$ GridWorld, shown in the inset, as a function of the number of trials. The same analysis is carried out for implementations with ideal (blue) and noisy (orange) phase-shifters. See Sec. \ref{sec_exp_imperfections} for details on how experimental imperfections were modeled. Curves are averaged over $10^4$ agents ($\lambda=8$, $\eta=0.11$ and damping $\gamma=0.999$ applied every 100 steps), while the gray band excludes lengths below the minimum (19 steps).
	Inset: Path taken by a single, noisy, random  agent after 150 trials. The green sphere ($\vec{p}_A=(3,1,4)$) and the blue sphere ($\vec{p}_{R}=(9,9,9)$) represent the agent and the reward, respectively, while blocks represent untraversable 3D walls.
	}
	\label{fig:4}
\end{figure}

The above formulation can be extended to more complex scenarios, which include higher-dimensional mazes with walls, sophisticated moves and/or penalties. For our investigation we employed a 3D GridWorld with walls: whenever the agent tries to move onto the border of the grid or onto a wall, a time step is counted but no movement occurs.
We chose a 3D maze, rather than a 2D or a 4D grid, to investigate more complex configurations that could still be visually inspected. As an example (see inset in Fig. \ref{fig:4}), we considered a $10 \times 10 \times 10$ GridWorld where the agent starts at position $\vec{p}_A=(3,1,4)$ and a reward is hidden at position $\vec{p}_{R}=(9,9,9)$.
Fig. \ref{fig:4}a shows the average learning curve numerically simulated for a photonic agent navigating this maze. We observe that the average path length rapidly decreases with the number of trials, from $\sim 10^3$ (where the agent behaves like a random walker) to values close to the minimum path length (19 in this case).

The same numerical analysis was carried out simulating a non-ideal implementation of photonic PS, to test to what extent experimental imperfections are expected to spoil the process. To this end, each time phases were adjusted in the simulated device, Gaussian noise was added on top of the ideal value (a more detailed description on how imperfections were modeled is reported in Sec. \ref{sec_exp_imperfections}). Remarkably, we find that a realistic amount of noise can even aid the learning process, a feature that can be ascribed to an enhanced tendency of the agent to explore new paths. In Sec. \ref{sec_exp_imperfections}, we also expand on this aspect, which is reminiscent of the phenomenon of stochastic resonance \cite{Note3}, providing a visual intuition in support of this interpretation.
Eventually, the fact that realistic levels of noise can enhance the agent's learning process makes the present approach even more appealing for a concrete implementation. Indeed, not only the architecture exhibits a natural resilience to noise, but also this very resilience relaxes the (often challenging) technological requirements for isolation and stability.


\section{\lowercase{t}-PS with generalization and abstraction}

\label{sec:multilayerPS}

While the two-layer and the tree-based implementations of PS have the same representational power (see Sec. \ref{sec:A_multilayerPS}), t-PS provides an additional structure that can be exploited to boost the learning process. As we will see, this feature allows an agent to exhibit simple forms of abstraction and generalization, which play a central role in artificial intelligence  \cite{Ponsen10}.
Abstraction is the ability of an agent to filter out less relevant details, a process that involves a modification in the representation of the object. Generalization corresponds to the ability to identify similarities between objects, without necessarily affecting their representation.
In this section, we will describe how an agent can take advantage of these features by suitably ordering the clips over the output modes according to some measure of relevance, such as the reward.

\subsection{Generalization and abstraction}
\label{sec:multilayerPS_generalization}

To introduce the notions of generalization and abstraction in the present architecture, let us start by considering the simplest case of a 2D GridWorld in the $XY$ plane without walls.
Given the tree structure of t-PS, we can expect there to be a beneficial arrangement of action clips over the outputs. Nodes in t-PS can represent meaningful sub-decisions towards a final decision made at the leaf nodes. Since nodes closer to the root are updated more regularly, sub-decisions can, in principle, be learned before the final policy is obtained. Of course, initially, nodes are not necessarily ordered in a way that has a meaningful interpretation. However, the agent can sort them during the learning process such that intermediate nodes obtain meaning which, in turn, guides the agent's decision-making.

\begin{figure}[t]
	\includegraphics[width=\linewidth]{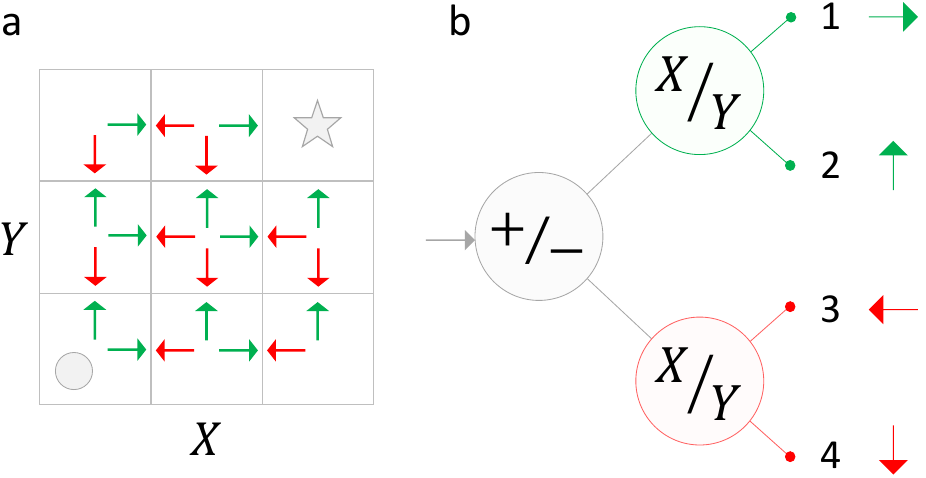}
	\caption{\textbf{Generalization in GridWorld.} t-PS can exploit symmetries in a task environment to boost the learning process. In a 2D GridWorld (a) where the agent (circle) and the reward (star) are initially located at opposite corners, we can associate with modes (1,2,3,4) the actions ($\rightarrow,\uparrow,\leftarrow, \downarrow$), so that the agent can learn to focus on the first two by adjusting just one parameter (b).}
	\label{fig:5}
\end{figure}

Motivated by the above considerations, we propose a simple mechanism, which we call \emph{defragmentation}, that is specifically designed to address this issue, though its benefits are not limited to this scenario. The name defragmentation is inspired by the usual process that occurs in hard-disks, which improves performance by reallocating fragments of memory according to dependencies and usage. The mechanism consists of (1) keeping track of the cumulative reward assigned to each action and (2) sorting actions over the output modes according to their respective cumulative reward. More sophisticated rules can also be designed for step (1), perhaps tailored to capture correlations in time or more intricate patterns between actions. From a practical perspective, step (2) only requires to compute the new phases that produce the reordered probability distribution (see Sec. \ref{sec:programming_binary_tree}). In any case, whenever there are two or more rewarded actions, this mechanism favors the separation between good and unfavorable actions. It is precisely in its capability of grouping together actions of comparable relevance, e.g. similar collected rewards in the present context, that the agent expresses an elementary form of generalization \cite{melnikov2017generalization}. 
For instance, in a 2D GridWorld actions can be conveniently organized according to a hierarchy of criteria (Fig. \ref{fig:5}), e.g. move 'forward' or 'backwards' and move 'along X' or 'along Y', resulting in composite actions such as 'up' ('forward' and 'Y') or 'left' ('backwards' and 'X').
Numerical analyses involving defragmentation on both 2D and 3D GridWorld show that the agent does autonomously discover structures analogous to the one in Fig. \ref{fig:5}, suggesting that this generalization feature is beneficial and informative, and that it can be used in more complex scenarios.

Naturally, defragmentation, as a way of knowledge exploitation, consumes time that has to be balanced with that reserved for exploration. Nevertheless, in the usual compromise between exploration and exploitation \cite{SUTTON90}, the longer the agent explores the environment to assess the quality of an action, the more its generalization process will be reliable and successful. 
At a certain time, once a stronger representation is built in its memory, the agent could even perform a sort of abstraction by cutting out the least relevant actions, so as to focus only on those that are deemed more favorable. In RL tasks with large-scale action spaces, this process could even be iterated to progressively reduce the search space for good actions. Indeed, the photonic architecture enables this mechanism to be straightforwardly implemented, by simply setting specific transition probabilities to 0 or 1, which isolates all the subsequent branches of optical components. This feature could, in turn, entail a reduction in computational resources and, possibly, in learning time.

\subsection{Exploiting the tree-like structure}
\label{sec:multilayerPS_bandit}

To provide quantitative evidence for the above considerations, we numerically applied defragmentation to another standard problem in RL, the multi-armed bandit \cite{SUTTON90}.
In its general formulation, an agent is presented with $N$ bandits (for instance, slot machines) characterized by a probabilistic reward function and, at each time step, the agent is allowed to pull the arm of one of the bandits (which issues a reward drawn from the corresponding distribution).
Effectively, this gives an environment with one state and $N$ possible actions.
We consider a variant of the problem with additional structure in its action space, referred to in the literature as \emph{combinatorial} multi-armed bandit \cite{Chen13}. In this task, bandits (i.e., actions) are grouped in sub-categories according to a set of features. In the example described above, these features could be the casino, city, country, etc. the slot machine is situated in. This structure is provided to the agent at an abstract level (the dependence between features is not specified)  by dividing the allowed actions into several sub-actions. As a result, the action space $A = \{1, ..., N\}$ factorizes to $A = A_1 \times A_2 \times ... \times A_k$, where $|A_i| = n_i$ is the number of possible choices for sub-action $A_i$, and $N = \prod_i n_i$. This kind of factorization is analogous to the decomposition of the state and action space into categories that was considered in Ref. \cite{melnikov2017generalization}, except that the structure we consider here is imposed on actions. For simplicity, let us assume that a deterministic reward $r_a$ is associated with each action $a=(a_1,...,a_k)$, but that this reward distribution depends (partially) on the structure of the action space. The agent can then exploit the factorized structure to choose the best sub-actions according to their influence on the reward. In this regard, the proposed architecture can be particularly effective since consecutive levels can separately focus on each $A_k$. Moreover, a mechanism to rearrange the layers (such as the defragmentation described in Sec. \ref{sec:multilayerPS_generalization}) can shift the layers associated with the most relevant sub-actions closer to the root, capturing correlations between actions and facilitating learning. In the above example, the agent could learn that the choice of a city is more relevant than the choice of a particular casino in that city, because casinos in a certain city are more lucrative, and choose the city earlier in the deliberation.

\begin{figure}[b]
	\includegraphics[width=\linewidth]{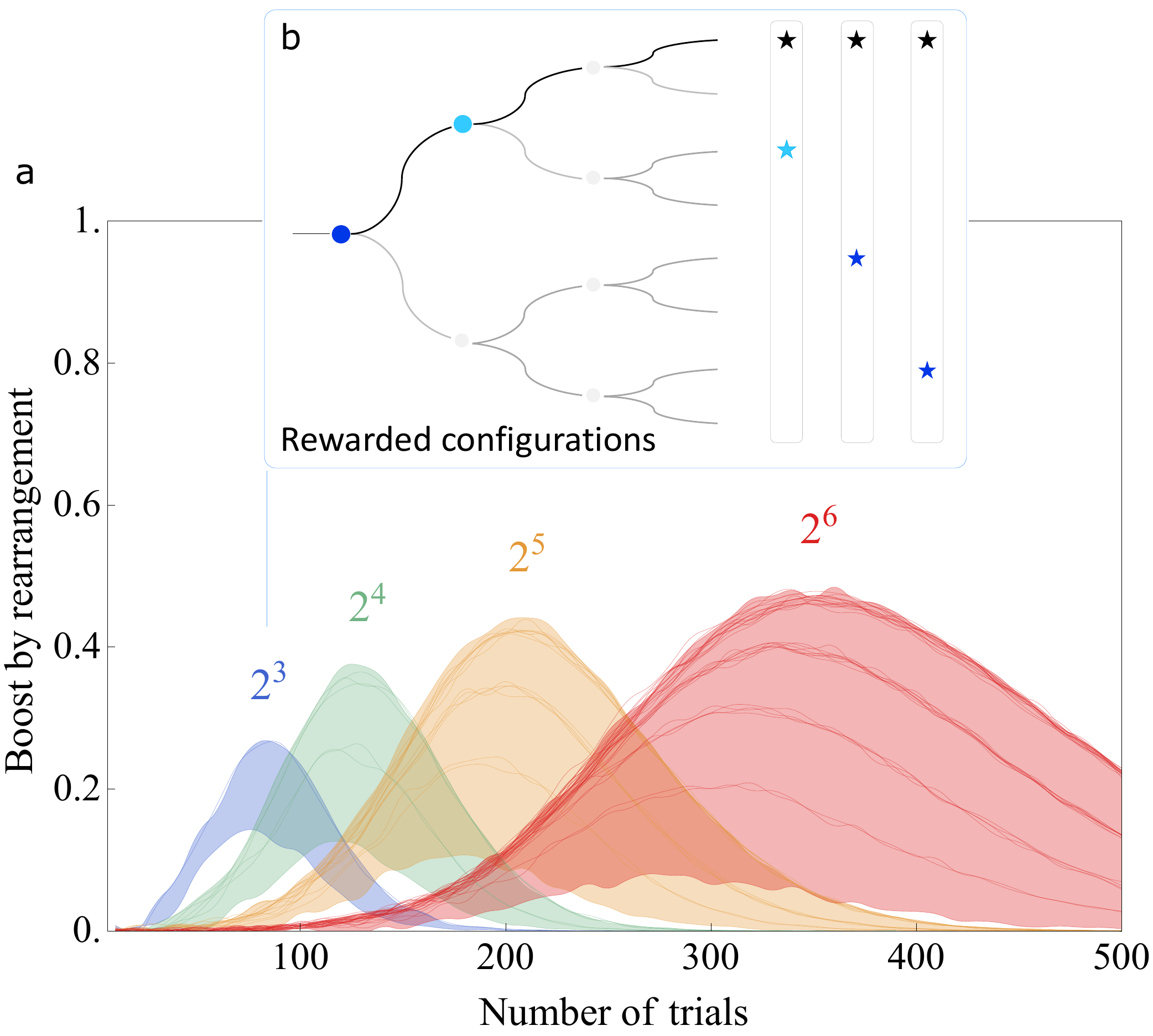}
	\caption{\textbf{Boosting the learning process in t-PS.} Learning  can be sped up in tasks with structured action spaces like the combinatorial multi-armed bandit \cite{Chen13}, by taking advantage of the tree-like structure. 
	a) Difference (\textit{boost}) between the average reward collected with and without defragmentation of the agent's memory, i.e. a dynamical rearrangement of the actions over the output modes. The analysis is carried out for actions spaces of size $2^d$, with $d=3, ..., 6$, with only two actions rewarded (b): one fixed on the first output mode, the other one displaced progressively further over the other modes. For each $d$, the magnitude of the boost depends on the number of layers (from 1 to $d-1$) where the rewarded paths differ: neighboring (faraway) modes lead to smaller (higher) boosts. For clarity, curves are interpolated connecting one point every 10. Averages are computed over $5 \times 10^3$ PS agents ($\lambda=0.025$, $\gamma=0.9975$).}
	\label{fig:6}
\end{figure}

We expand on the above considerations in more detail in Sec. \ref{sec:factorized_spaces} with a simple example. In the following, we will focus on the performance boost induced by the defragmentation of the action space. Fig. \ref{fig:6} shows quantitative evidence of this boost in an instance of the bandit problem where two actions are always rewarded.
%
%
Analogous advantages can be found in the 3D GridWorld described in Sec. \ref{sec:TestingGridWorld}, where only a subset of directions is relevant and grouping them is beneficial for the agent. 
In particular, these numerical results show that defragmentation allows to speed up the learning process, i.e. fewer trials are required to find an optimal policy. This situation is indeed typical in RL, where exploitation of current knowledge allows to reduce the time spent on exploration. From a practical perspective, this feature facilitates learning scenarios where interactions with the environment are costly.  
For these reasons, the proposed t-PS appears as a promising platform in the framework of RL, being able to support key features for artificial intelligence (in the form of a basic generalization and abstraction) while preserving a good control over its operation and performance.


\section{Discussion}

The development of autonomous agents capable of learning by interacting with an environment has seen a tremendous surge of interest over the past decade \cite{Mnih2015, Silver2016, openAI, Arulkumaran19}. Recently, RL has even claimed its place in the list of the top breakthrough technologies with the largest and broadest impact \cite{MITreviewRL}.
Similarly, the design of neuromorphic application-specific hardware has attracted massive attention due to its enhanced computational capabilities in terms of speed and energy efficiency \cite{Thakur18}. In this work, we propose a blueprint for an application-specific integrated photonic architecture capable of solving problems in RL.
Within this framework, the architecture easily accommodates various well-established RL algorithms such as SARSA, Q-learning, and PS.
Also, its simple and scalable design warrants near-term implementations and is apt for embedding in portable devices. Indeed, all required optical components have already been experimentally demonstrated on integrated circuits \cite{Sun15, Komljenovic16, Flamini_2018, Atabaki2018, Harris2018, Pérez2018}.

We investigated the proposed platform both numerically and analytically, confirming the efficacy of the model also under realistic, imperfect experimental conditions.
Besides its efficacy, the architecture enables a novel implementation of PS (t-PS) that is inspired by the geometry of the integrated circuit. This model does not only exhibit some key features of artificial intelligence, namely generalization and abstraction, but can also boost its learning performance via autonomous defragmentation of its memory.
Indeed, both numerical and analytical results suggest that t-PS performs at least as well as the simulated standard PS model, which has already found various applications \cite{simon2016,melnikov2017active,poulsennautrup2018optimizing,Ried19}.
Eventually, we envisage the experimental realization of a photonic RL agent which successfully exploits all these features within an optical environment.

\section*{Acknowledgements}

This project has received funding from the European Union's Horizon 2020 research and innovation programme under the Marie Skłodowska-Curie grant agreement No 801110 and the Austrian Federal Ministry of Education, Science and Research (BMBWF). It reflects only the author's view and the Agency is not responsible for any use that may be made of the information it contains.
HPN, LMT, SJ, and HJB acknowledge support from the Austrian Science Fund (FWF) through the projects DK-ALM:W1259-N27 and SFB BeyondC F71. HJB was also supported by the Ministerium f\"ur Wissenschaft, Forschung, und Kunst BadenW\"urttemberg (AZ:33-7533.-30-10/41/1).


\section{Appendix}


\subsection{Programming the architecture}
\label{sec:programming_binary_tree}

Here we describe how to create an arbitrary output probability distribution in t-PS by tuning the parameters available in a photonic architecture, i.e. $\theta$.
Given a clip $c$ and an associated output probability distribution $\{q_{c,i}\}$, we can analytically retrieve the set of phases $\theta_{c,kl}$ that reproduces the probability distribution in the $n$-layer tree architecture.
To this end, we consider the ratio $\xi_{c,kl}$ of the probabilities of taking the upper ($\hat{p}_{c,kl}$) or the lower paths at node ($k$,$l$) (Fig. \ref{fig:2})
\begin{equation}
	\xi_{c,kl} = \frac{ \hat{p}_{c,kl}  }{ 1-\hat{p}_{c,kl}  } =
	\frac{ \sum\limits_{i\in \mathcal{U}_{kl}}  q_{c,i} }{\sum\limits_{i\in \mathcal{D}_{kl}} q_{c,i} }
\label{eq:mathematica_formula}
\end{equation}
\noindent  where $k \in [1,n]$, $l \in [1,2^{k-1}] $ and the sum in the numerator (denominator) runs over the output modes associated with the upper (lower) path. In particular, if we label the output nodes from 1 to $2^n$, we find that
\begin{equation}
    \begin{array}{c}
		\mathcal{U}_{kl} = [1+(2l-2)\,2^{n-k}, \, (2l-1)\,2^{n-k} ]   \\
		\ \\
		\mathcal{D}_{kl} = [1+(2l-1)\,2^{n-k}, \, 2l \, 2^{n-k} ]   \\
    \end{array}
\label{eq:intervals}
\end{equation}
Writing $\hat{p}_{c,kl} = \sin^2 \theta_{c,kl}$ (as we are dealing with phases in the photonic t-PS), we finally get $\theta_{c,kl} = \arctan \sqrt{ \xi_{c,kl} }  $.


\subsection{Update rules for t-PS}
\label{sec:A_multilayerPS}

 In this section, we discuss three relevant rules to update phases in the photonic t-PS architecture. The section is structured as follows: first, we derive the rule of Eq. \ref{eq:photonic_map}. Then, we show that this architecture can also simulate the behavior of a two-layered PS where probability distributions are either calculated from normalized $h$-values (Sec. \ref{sec:2Lstandard}) or the softmax function of $h$-values (\ref{sec:2Lsoftmax}). Hence, t-PS can reproduce the results reported in the literature on PS.


\subsubsection{Derivation of $\theta_{\chi}$ in Eq. \ref{eq:photonic_map}}
\label{sec:multilayerSoftmax_update}

Below we derive an expression for $\phi(\chi_{c,kl})$, which appears in Eq. \ref{eq:photonic_map} through the mapping $\theta(\chi_{c,kl}) = \frac{\pi}{4} + \phi(\chi_{c,kl})$, such that $\hat{p}_{c,kl} = \sin^2\left(\theta(\chi_{c,kl})\right)$ yields an approximation of the softmax of the $h$-values at node ($k$,$l$).
For now, let us omit the $\chi$-dependency and consider a quantity $\phi_{c,kl}$. In this scenario, the probability $\hat{p}_{c,kl}$ of taking the upper path ($\uparrow$) is
\begin{equation}
\hat{p}_{c,kl}  =  \frac{e^{h_{c,kl}^{\uparrow}}}{e^{h_{c,kl}^{\uparrow}}+e^{h_{c,kl}^{\downarrow}}} 
=    \frac{1}{2} + \frac{1}{2} \tanh \frac{\Delta h_{c,kl}}{2}
\label{eq:PS_ml}
\end{equation}
\noindent  where we used $\Delta h_{c,kl} = h_{c,kl}^{\uparrow} - h_{c,kl}^{\downarrow} $ and $e^{2x} = \frac{1+\tanh{x}}{1-\tanh{x}}$.
The same scattering probability in a photonic circuit can be conveniently written as
\begin{equation}
\hat{p}_{c,kl}    \equiv  \sin^2 \left( \frac{\pi}{4} + \phi_{c,kl} \right) =   \frac{1}{2} + \frac{1}{2} \sin \left(   2 \phi_{c,kl}  \right)
\label{eq:}
\end{equation}
\noindent  which leaves us with the identity 
\begin{equation}
\phi_{c,kl}  = \frac{1}{2} \arcsin \left(   \tanh \left( \frac{\Delta h_{c,kl}}{2} \right)  \right) .
\label{eq:arcsin_map}
\end{equation}
\noindent  Finally, observing that $\arcsin \left( \tanh (x)\right) \approx \frac{\pi}{2} \tanh (\frac{2x}{\pi}) \; \forall x \in \mathbb R $, we can approximate Eq. \ref{eq:arcsin_map}  as  
$ \phi_{c,kl} = \frac{\pi}{4} \tanh (  \frac{\Delta h_{c,kl}}{\pi} ) $, which leads to Eq. \ref{eq:photonic_map} when $\chi_{c,kl} \equiv \frac{\Delta h_{c,kl}}{\pi} $ and $\theta(\chi_{c,kl}) = \frac{\pi}{4} + \phi(\chi_{c,kl}) $.


\subsubsection{Reproducing two-layer PS with standard probabilities}
\label{sec:2Lstandard}

We look for an update rule $\theta_{c,kl} \mapsto f(\theta_{c,kl})$ in t-PS that reproduces the update on the standard probabilities $q_{c,i}=h_{c,i} / \sum_j h_{c,j}$ of the two-layer PS \cite{briegel2012projective}.
Clearly, t-PS can, in principle, reproduce the probabilities in the 2-layered PS since it can reproduce any probability distribution, as we showed in Sec. \ref{sec:programming_binary_tree}. However, it is not obvious that there exists an update rule on the parameters $\{\theta\}$ that simulates an update on the $h$-values in the 2-layered PS. Therefore, we will first show that (i) there exists a local update rule $g(\cdot)$ on  $\{\hat{p}^{(t)}_{c,kl}\}$ such that  $\{g(\hat{p}^{(t)}_{c,kl})\}$ represents $ \{q^{(t+1)}_{c,i}\} \forall t$.
Then, ($ii$) we will express $\hat{p}_{c,kl}$ using $\theta_{c,kl}$, which gives the desired update rule $\theta_{c,kl} \mapsto f(\theta_{c,kl})$. 
For brevity, in the following we ignore the time index because it suffices to consider a single update.

($i$) We start by considering the ratio $\xi_{c,kl}$ of the transition probabilities at node ($k$,$l$)
\begin{equation}
		\xi_{c,kl} =\frac{ \hat{p}_{c,kl}  }{ 1-\hat{p}_{c,kl}  } =
		\frac{ \sum\limits_{i \in \mathcal{U}_{kl}}  h_{c,i} }{ \sum\limits_{s \in \mathcal{B}_{kl}} h_{c,s} }
		\frac{ \sum\limits_{s \in \mathcal{B}_{kl}} h_{c,s} }{\sum\limits_{i \in \mathcal{D}_{kl}} h_{c,i} } =
		\frac{ u_{c,kl} }{ d_{c,kl} } 
\label{eq:start}
\end{equation}
\noindent where $\mathcal{U}_{kl}$ and $\mathcal{D}_{kl}$ are defined in Eq. \ref{eq:intervals},  $\mathcal{B}_{kl} = \mathcal{U}_{kl}\cup \mathcal{D}_{kl}$ is the set of branch indexes associated with all output modes reachable from node ($k$,$l$) and
\begin{equation}
		u_{c,kl} = \sum\limits_{i \in \mathcal{U}_{kl}}  h_{c,i}  \qquad \quad 
		d_{c,kl} = \sum\limits_{i \in \mathcal{D}_{kl}}  h_{c,i}
\label{eq:u_d}
\end{equation}
\noindent  Since Eq. \ref{eq:start} holds at each time step, when the transition to a certain clip $c'$ is rewarded (by a value $\lambda_{c'}$, i.e. $h'_{c,c'} = h_{c,c'} + \lambda_{c'}$) we have
\begin{equation}
\xi'_{c,kl} = \left\{
    \begin{array}{ll}
		\frac{  u_{c,kl} }{  d_{c,kl} } + \frac{ \lambda_{c'}}{  d_{c,kl} }   &\quad c' \in \mathcal{U}_{kl} \\
	\ \\
		\frac{  u_{c,kl} }{  d_{c,kl} } \left( \frac{1}{1+ \frac{ \lambda_{c'}}{  d_{c,kl} } } \right)   & \quad c' \in \mathcal{D}_{kl} \\
    \end{array}
     \right. 
\label{eq:2L_standard_paths}
\end{equation}
\noindent depending on whether the rewarded action is related to the upper path or to the lower path.
Defining $ N _{c,kl} = u_{c,kl}  + d_{c,kl}  $ we obtain
\begin{ceqn}
	\begin{align}
	\left\{
         \begin{array}{ll}
        \  \xi_{c,kl} ' = \xi_{c,kl} \Bigg[ 1 +  \left(1+ (\xi_{c,kl}) ^{(-1)^{c'_k+1}}   \right) \frac{\lambda_{c'}}{ N_{c,kl}}    \Bigg] ^{(-1)^{c'_k}}    \\
		\ \\
                  \   N_{c,kl}'  = N_{c,kl} + \lambda_{c'}   \\
                \end{array}
                           \right.            
	\label{eq:update_flat_linear}                        
	\end{align}
\end{ceqn}
\noindent  where $c_k'=0$ ($c_k'=1$) if $c' \in \mathcal{U}_{kl}$ ($c' \in \mathcal{D}_{kl}$),  and with the initial settings $\, \xi_{c,kl}^{(t=0)} = 1$ and $\, N _{c,kl}^{(t=0)} = 2^{n-k+1}$. Indeed, $c_k'$ can be seen as the $k$th digit of $c'$ written in base 2 ($c'_k = 0$ for upper paths, $c'_k = 1$ for lower ones). Eq. \ref{eq:update_flat_linear} shows that there exists an update $g(\cdot)$ on $\{p_{c,kl}\}$ that reproduces the update on $\{q_{c,i}\}$ in the two-layered PS.

($ii$) By inserting $\hat{p}_{c,kl} = \sin^2 \theta_{c,kl}$ into Eq. \ref{eq:start} we obtain $\theta_{c,kl} = \arctan \sqrt{ \xi_{c,kl} } $. This connection allows the reformulation of Eq. \ref{eq:update_flat_linear} in terms of $ \theta_{c,kl}$, which gives the update rule we were looking for to reproduce the two-layer PS in t-PS.


\subsubsection{Reproducing two-layer PS with softmax function}

\label{sec:2Lsoftmax}

We now describe an update rule on t-PS that simulates the two-layer PS with softmax function. The softmax function (see Eq. \ref{eq:PS_prob_softmax}) is a convenient tool to construct a probability distribution $\{p_i\}$ from a set of non-normalized quantities $\{h_i\}$.
The derivation of the update rule in this case develops in the same manner as in Sec. \ref{sec:2Lstandard}, through two main steps.

($i$) Considering the softmax function and the update rule $ h'_{c,c'} = h_{c,c'} +  \lambda_{c'}$, following Eq. \ref{eq:start} and Eq. \ref{eq:u_d} we have
\begin{equation}
	\xi_{c,kl} = \frac{ p_{c,kl} }{  1-p_{c,kl} } =
	\frac{ \sum\limits_{i\in \mathcal{U}_{kl}} e^{\beta h_{c,i} }}{ \sum\limits_{i\in \mathcal{D}_{kl}} e^{\beta h_{c,i}}  } =
	\frac{  U_{c,kl} }{ D_{c,kl} }
\label{eq:ratio_sigmas_softmax}
\end{equation}
Depending on the rewarded path as in Eq. \ref{eq:2L_standard_paths}, we get 
\begin{ceqn}
	\begin{align}
	\xi_{c,kl}' = 
	\left\{
        \begin{array}{ll}
			\xi_{c,kl} + \left( e^{\beta \lambda} -1 \right) \frac{ e^{\beta h_{c,c'}}}{ D_{c,kl}  }    & \, c' \in \mathcal{U}_{kl} \\
		\ \\
			\xi_{c,kl}\left( 1 + \left( e^{\beta \lambda} -1 \right) \frac{ e^{\beta h_{c,c'}}}{ D_{c,kl}  }  \right)^{-1}   & \, c' \in \mathcal{D}_{kl} \\
        \end{array}
            \right.            
	\label{two_paths_softmax}                        
	\end{align}
\end{ceqn}
Ideally, we would like an update rule that involves only the quantity that is being updated, i.e. $ \xi $. In order to express the ratio $  e^{\beta h_{c,c'}} /  D_{c,kl}  $ in terms of $\xi$, we first observe that 
\begin{equation}
		\frac{ e^{\beta h_{c,c'}}}{ D_{c,kl}  } = \frac{ e^{\beta h_{c,c'}}}{U_{c,kl}+ D_{c,kl}} \frac{U_{c,kl} + D_{c,kl}}{ D_{c,kl}  } = \frac{ p(c'|kl) }{ 1-\hat{p}_{c,kl} }
\end{equation}
\noindent  where
$p(c'|kl)$ is the probability of a photon exiting the output mode corresponding to the next rewarded clip $c'$ given that it is at node ($k$,$l$)
\begin{equation}
		p(c'|kl) = \prod\limits_{(v,w) \in \Gamma_{kl,c'}} \left( \hat{p}_{c,vw}^{1-c_v'} \, (1-\hat{p}_{c,vw})^{c_v'}  \right)
\end{equation}
\noindent  evaluating the product over all nodes $\{(v,w)\}$ that connect $(k,l)$ to $c'$ in the binary tree. Since $\hat{p}_{c,kl} = \frac{ \xi_{c,kl} }{ 1 + \xi_{c,kl} }$,
%
%
we can finally express Eq. (\ref{two_paths_softmax}) as follows
\begin{equation}
    \xi_{c,kl}' = \xi_{c,kl}  \Bigg[ 1 + (e^{\beta \lambda}- 1)  \prod\limits_{(v,w) \in \Gamma_{kl,c'} }     
    \frac{{\xi_{c,vw}}^{c_v'}}{1+\xi_{c,vw} }  \Bigg] ^{(-1)^{c'_k}}
\label{update_2L_softmax}
\end{equation}
Note that this expression only involves the quantity $\xi$.

($ii$) Using again $\theta_{c,kl} = \arctan \sqrt{ \xi_{c,kl}}$, Eq. \ref{update_2L_softmax} provides the update rule to simulate the two-layer PS with softmax function in the t-PS architecture.


\subsection{Processing time and energy consumption}
\label{sec_scaling}

In this section, we discuss the computational complexity of the proposed photonic platform and of an ideal application-specific integrated circuit (ASIC), employing sampling algorithms and data structures which are best suited for the present application.
To this end, let us assume that both the photonic hardware and the ASIC store weights, i.e. $\chi_{c,kl}$ and $h_{ij}$ respectively, in an on-board memory, ideally a cache. Both architectures must perform three computational tasks: (i) updating and (ii) preprocessing  $N$ weights, and (iii) sampling from preprocessed data. Let us discuss each part in order.

(i) In the photonic architecture, updating the in-memory weights requires adjusting $\log N$ $\chi$-values along a path in the binary tree of Fig. \ref{fig:2}. Basically, this operation corresponds to $\mathcal{O}(\log N)$ number of FLOPS. Similarly, we only update a single $h$-value in the ASIC. However, in order to make steps (ii) and (iii) efficient, we demand that the $h$-values are ordered in a sorted list. Then, a single update may very well disturb this sorting and require up to $\mathcal{O}(N)$ operations to recover from. Therefore, we assume that the $h$-values are stored in a self-balancing tree data structure, a so-called B-tree~\cite{Comer1979}. This data structure not only allows easy access in $\mathcal{O}(\log N)$ computational time but also includes insertion and deletion operations that maintain the order of elements while requiring the same logarithmic time complexity.

(ii) Preprocessing in the photonic architecture requires adjusting $\log(N)$ PCM phase-shifters by evaluating $\theta(\chi)$ for the updated values, each requiring $\sim 10^2$ pJ at the nanosecond scale \cite{Rios19}. This is comparable to the power consumption of ideal, specialized computing devices at $\sim 1\textrm{pJ}/\textrm{FLOP}$~\cite{Shen17} and may be improved due to the broad applicability of PCMs for energy storage, information processing, and optical communication \cite{Wuttig17, Miller18}. For comparison, a general-purpose computing device requires $\sim 1$ nJ per DRAM access and $\sim 10$ pJ per cache access~\cite{Horowitz14}. In the ASIC, we prepare for sampling by creating auxiliary data from the sorted list of weights, in accordance with the preprocessing outlined in the \textsc{SortedProportionalSampling} algorithm proposed in Ref.~\cite{Bringmann2017}. This preprocessing requires $\mathcal{O}(\log^2 N)$ computational time when data are stored as a B-tree. 

(iii) In both cases sampling takes constant time: in the photonic device, sampling reduces to the generation and detection of a single photon, while the query complexity of \textsc{SortedProportionalSampling} is $\mathcal{O}(1)$ once preprocessing is concluded \cite{Bringmann2017}.

In summary, both the photonic architecture presented in the main text and the ASIC described here have about the same computational complexity $\mathcal{O}(\log N)$ \cite{Horowitz14}. Note that, in principle, we need to take into account both memory access operations and FLOPS when estimating the energy cost. However, assuming a highly localized architecture approximately equalizes the power consumption of memory accesses and FLOPS.


\subsection{Role of experimental imperfections}
\label{sec_exp_imperfections}

Experimental noise and fabrication imperfections represent an unavoidable issue for any implementation. Their detrimental effects on device fidelities can also increase rapidly for applications that involve multiphoton interference in large-size interferometers \cite{Flamini17, Burgwal17, Russell17}. As we discuss below, however, the tolerance to noise in the proposed architecture is comparatively high for at least two reasons.
(i) The approach described in this work involves only single-photon evolutions in linear-optical circuits, reducing the influence of unbalanced phases that is critical for multiphoton interference. Also, the circuit depth scales logarithmically with the number of modes, thus limiting propagation losses.
(ii) The additional randomness induced by noise can play a positive role in the operation of the device. In fact, since decision-making consists of single-photon random walks, random deviations from the ideal probability distributions lead to a tendency to explore alternative paths, without sticking to the estimated policy (as opposed to greedy approaches).

\begin{figure*}[t]
	\includegraphics[width=0.9\linewidth]{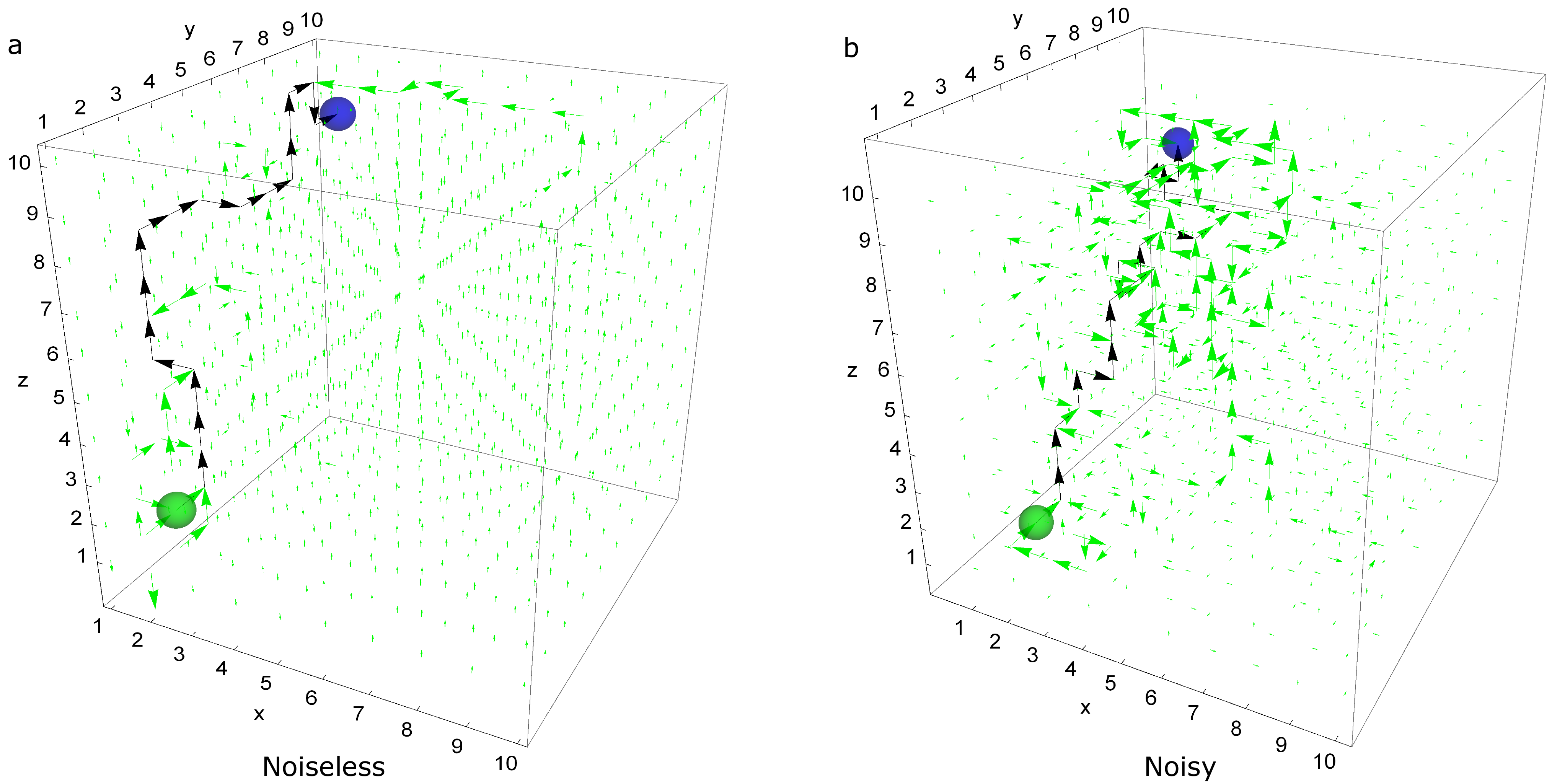}
	\caption{\textbf{Noise-enhanced exploration in GridWorld.} Noise in the photonic implementation represents an additional source of randomness in the learning process, which enhances the likelihood of exploring new paths, as well as avoid getting stuck with suboptimal behavior. This figure shows a comparison between the PS policies learned (a) without noise or (b) with noise: noisy plots tend to exhibit larger clouds, meaning that more paths have been explored and reinforced in the same time.
	Here, the green sphere ($\vec{p}_A=(2,2,2)$) and the blue sphere ($\vec{p}_{R}=(2,9,9)$) represent the PS agent and the reward, respectively. The learning parameters are $\lambda=8$ and $\eta=0.11$, and damping with $\gamma=0.999$ is applied every 100 steps as in Fig. \ref{fig:3}.
	Green arrows describe the most probable action the agent would take in each cell, with a size proportional to the probability. Black arrows highlight a single path taken by the agent after the learning process. }
	\label{fig:7}
\end{figure*}

To investigate this aspect, in Fig. \ref{fig:7} we consider a noisy architecture used to solve a 3D GridWorld analogous to Fig. \ref{fig:4}.
%
%
To model noise, we follow a standard approach for tunable photonic circuits, where each beamsplitter $U_{c,kl}^{BS}$ is physically implemented as a Mach-Zehnder interferometer with a tunable phase-shifter between two symmetric beamsplitters
\begin{ceqn}
	\begin{align}
		U_{c,kl}^{BS} = \frac{1}{2}
			\begin{pmatrix} 
				 1 & -1  \\
				 1 & 1  \\
			\end{pmatrix}
			\begin{pmatrix} 
				 1 &  0  \\
				 0 & e^{\imath \, \theta_{c,kl}} \\
			\end{pmatrix}
			\begin{pmatrix} 
				 1 & -1  \\
				 1 & 1  \\
			\end{pmatrix}			
	\end{align}
\end{ceqn}
Gaussian noise is then added to the phases $\theta_{c,kl}$, to simulate imperfect settings or mechanical instabilities. Specifically, in Fig. \ref{fig:7} we assume ideal beamsplitters transmissivities to isolate the contribution of phase errors, however a similar behavior is observed when noisy beamsplitters are considered. We observe that noise-free implementations (Fig. \ref{fig:7}a) tend to remember only very few very good paths in the agent's memory. Conversely, noisy implementations (Fig. \ref{fig:7}b) tend to explore many more effective paths, eventually giving rise to a cloud of paths that connect to the reward from different locations.
We emphasize that, even though the plot only displays the behavior of a single agent on a single maze, the above results were found to hold for practically all the agents inspected. Numerical evidence for this advantage is provided in Fig. \ref{fig:4}, which shows that realistic levels of noise can indeed speed up the learning process.


\subsection{t-PS in factorized action spaces}

\label{sec:factorized_spaces}

\begin{figure*}[t!]
	\includegraphics[width=0.925\linewidth]{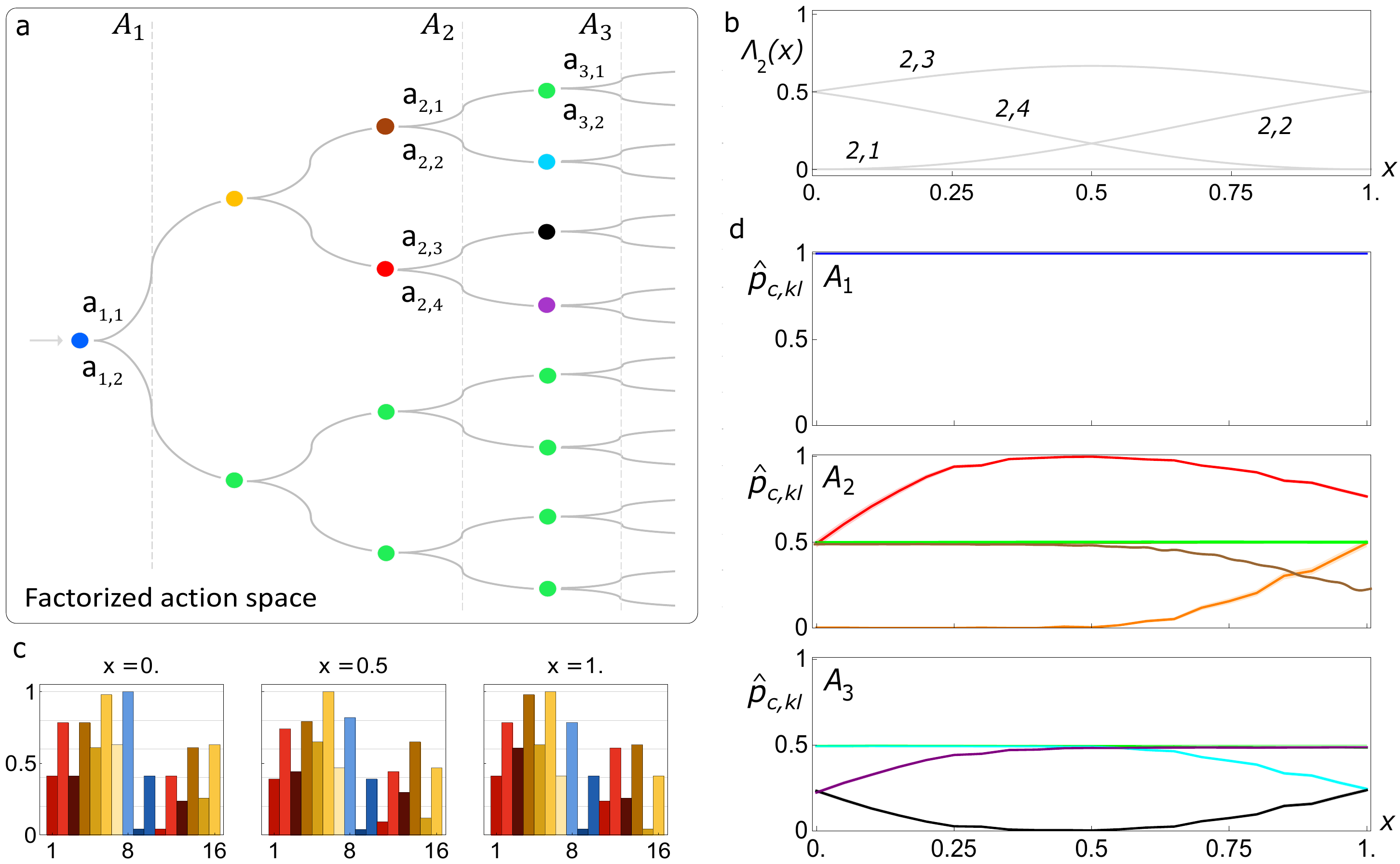}
	\caption{\textbf{Operation of the agent's memory in factorized problems.}  In each layer, an independent policy can be learned for a factorized action space, which allows the agent to boost the learning process. See text in Sec. \ref{sec:factorized_spaces} for a description of the example shown here.
	 a) Connection between beamsplitters and action subspaces within t-PS. Beamsplitters branches are labeled as $(i,a)$, where $i$ numbers the subspaces and $a$ the actions, and colors are used to link them to panel (d).
	 b) Evolution of the rewards $\Lambda_2(x)/\epsilon=(2-2x+2x^2)^{-1}(0,x^2,1,(1-x)^2)$ associated with subspace $A_2$ as a function of a parameter $x$. Labels $(2,a)$, with $a=1, ..., 4$, link the curves to the four actions of $A_2$.
	 c) Evolution of the full landscape of 16 rewards, here normalized to the maximum values. Separate colors are used to follow the evolution of each bar, and are not connected with the color scheme in panels (a) and (d).
	 d) Corresponding evolution of the probability $\hat{p}_{c,kl}$ of taking the upper path at each node ($k$,$l$). Curves are colored according to the layout in panel (a). Beamsplitters that do not change are shown in green. Values are averaged over $10^3$ agents, after $3\times10^3$ trials and rescaling the rewards by $\epsilon=0.004$. Clearly, the frequency with which beamsplitters in $A_3$ are traversed depends on the reward distribution $\Lambda_{2}(x)$. The reason why $\hat{p}_{c,kl} \ne 0$ for the first four beamsplitters in $A_3$, even though the action space is factorized, is that we are reporting only their average of their values, which oscillate between 0 (as expected) and 0.5 (when agents take other paths and beamsplitters are not enforced to change).
	 }
	\label{fig:A8}
\end{figure*}

The tree structure of t-PS is particularly convenient for problems with factorized action spaces. This is due to its architecture being able to capture the hierarchical structure of a problem, namely the correlation between different action subspaces. In Sec. \ref{sec:multilayerPS_generalization}, we discussed how defragmentation of the agent’s memory, which consists in reordering the way actions are assigned to the output modes, could allow forms of generalization and abstraction. In this section, we will show how defragmentation can capture the absence of correlation between action subspaces.
Specifically, we take a closer look at the internal operation of a simulated photonic agent, which is challenged to learn the optimal policy in an instance of the multi-armed bandit problem with independent action spaces \cite{SUTTON90}.
Let us consider a problem with three sub-actions associated with the spaces ($A_1$, $A_2$, $A_3$) of size (2, 4, 2), i.e. $A_1=(a_{1,1},a_{1,2})$, $A_2=(a_{2,1},a_{2,2},a_{2,3},a_{2,4})$, $A_3=(a_{3,1},a_{3,2})$, for a total of 16 actions (Fig. \ref{fig:A8}a). This construction is not natural in the formulation of the problem presented in Sec. \ref{sec:multilayerPS_bandit} (casino, country, ...), since we assume full independence between the components of the actions.
To investigate the dynamics of the internal settings, let us label the output modes ($m_1, ..., m_{16}$) by the action (or node) sequence, i.e. $m_1=(a_{1,1}, a_{2,1}, a_{3,1})$, $m_2=(a_{1,1}, a_{2,1}, a_{3,2})$ until $m_{16}=(a_{1,2}, a_{2,4}, a_{3,2})$. Also, let us assign rewards to the subspaces according to $\Lambda_1= \epsilon (0.95,0.05)$, $\Lambda_2(x)=\epsilon(2 - 2 x + 2 x^2)^{-1}(0,x^2,1,(1-x)^2)$ and $\Lambda_3=\epsilon(0.05,0.95)$ (Fig. \ref{fig:A8}b), $x$ and $\epsilon$ being a variable parameter and a rescaling factor, respectively. In this scenario, the beamsplitters in the first and last layer respectively control the behavior of $A_1$ and $A_3$, while those in the intermediate layers control $A_2$. Hence, we can monitor all the beamsplitters' transmissivities as rewards change with $x$ (Fig. \ref{fig:A8}c). As we show in Fig. \ref{fig:A8}d, the probability $\hat{p}_{c,kl}$ of taking the upper path at each node resembles the shape of $\Lambda_2(x)$ in Fig. \ref{fig:A8}b, in particular $\hat{p}_{c,kl} = 1$ for $A_1$ since the agent learns to make the first action no matter the value of $x$. Overall, it is possible to visually relate the curves in Fig. \ref{fig:A8}d to the underlying conditions described in Fig. \ref{fig:A8}c and (using colors that match curves and beamsplitters) in Fig. \ref{fig:A8}a. Furthermore, the fact that almost half of the beamsplitters (green) are not updated ($\hat{p}_{c,kl} = 0.5$, since their behavior is not relevant), can be seen as a form of abstraction that naturally occurred in the agent's memory.
Eventually, this connection between factorized actions spaces and internal parameters encourages to devise further mechanisms to enhance the learning process, which could simplify tasks in factorized (or factorizable) problems of higher dimensionality.


\section*{References}
\bibliographystyle{unsrt}

\end{document}